\documentclass[12pt,tightenlines,eqsecnum,floats,showpacs,nofootinbib,amsmath,amssymb,aps,prd]{revtex4}
\usepackage{graphicx,verbatim}
\usepackage{amsmath}
\usepackage{amsfonts}
\usepackage{amssymb}
\usepackage[colorinlistoftodos]{todonotes}
\usepackage{epstopdf}
\def\del{\partial}

\def\scri{\mathcal{I}}
\def\scrip{\mathcal{I}^{+}}
\def\rmd{\mathrm{d}}
\def\={\hat{=}}

\def\R{\mathcal{R}}

\def\F{\mathcal{F}}
\def\e{\mathfrak{e}}
\def\H{\mathcal{H}}
\newcommand{\pb}[1]{\hbox{\lower0.5ex\hbox{${}_{\leftarrow}$}}\kern-1.9ex{#1}}

\def\man{\mathcal{M}}
\def\M{\Sigma}

\def\Gp{\mathfrak{G}_{\rm Poin}}
\def\gp{\mathfrak{g}_{\rm Poin}}
\def\Diff{{\rm Diff}(\mathcal{I})}
\def\B{\mathcal{B}}
\def\K{\mathcal{K}}

\def\o{\omega}

\def\oM{\omega_{{\rm Max}}}

\def\Bcal{\mathcal{B}}
\def\Ecal{\mathcal{E}}

\def\ub{\underline}
\def\vx{\vec{x}}
\def\vk{\vec{k}}

\def\be{\begin{equation}}
\def\ee{\end{equation}}
\def\ba{\begin{eqnarray}}
\def\ea{\end{eqnarray}}

\def\f{\frac}

\def\qo{\mathring{q}}
\def\go{\mathring{g}}

\def\Do{\mathring{D}}
\def\Ho{\mathring{\mathcal{H}}}

\def\boxo{\mathring{\Box}}
\def\nablao{{\mathring{\nabla}}}
\def\ho{\mathring{h}}
\def\pso{\mathring{\Gamma}_{\rm cov}}
\def\omegao{\mathring{\omega}}

\def\ps{\Gamma_{\!\rm Cov}}
\def\psM{\Gamma_{\!\rm Cov}^{\rm Max}}

\begin{document}

\title{Asymptotics with a positive cosmological constant:\\ II. Linear fields on de Sitter space-time} 
\author{Abhay Ashtekar}
\email{ashtekar@gravity.psu.edu} %\affiliation{Institute for
%Gravitation and the Cosmos \& Physics
%  Department, Penn State, University Park, PA 16802, U.S.A.}
\author{B\'eatrice Bonga}
\email{bpb165@psu.edu} %\affiliation{Institute for Gravitation and
%the Cosmos \& Physics
%  Department, Penn State, University Park, PA 16802, U.S.A.}
\author{Aruna Kesavan}
\email{aok5232@psu.edu} \affiliation{Institute for Gravitation and the
Cosmos \& Physics Department, Penn State, University Park, PA 16802,
U.S.A.}

\begin{abstract}

Linearized gravitational waves in de Sitter space-time are analyzed  in detail to obtain guidance for constructing the theory of gravitational radiation in presence of a positive cosmological constant in { full, nonlinear general relativity}. Specifically: i) In the exact theory, the intrinsic geometry of $\scri$ is often assumed to be conformally flat in order to reduce the asymptotic symmetry group from $\Diff$ to the de Sitter group. Our {results show explicitly} that this condition is physically unreasonable; ii) We obtain expressions of energy-momentum and angular momentum fluxes carried by gravitational waves in terms of fields defined at $\scrip$; iii) We argue that, although energy of linearized gravitational waves can be arbitrarily negative in general, gravitational waves emitted by physically reasonable sources carry positive energy; and, finally iv) We demonstrate that the flux formulas reduce to the familiar ones in Minkowski space-time in spite of the fact that the limit $\Lambda \to 0$ is discontinuous (since, in particular, $\scri$ changes its space-like character to null in the limit).

\end{abstract}

\maketitle

\section{Introduction}
\label{s1}

A rich theory of isolated gravitating systems, developed systematically since the 1960s \cite{bondi, sachs1,sachs2,rp}, lies at the foundation of a large fraction of research in general relativity with zero cosmological constant. Examples include the gravitational radiation theory, classical and quantum aspects of black holes, and several major initiatives in geometrical analysis (see, e.g., \cite{abk1} for a summary). But observations strongly indicate that the cosmological constant is \emph{positive} in the universe we inhabit \cite{sndata}. Therefore it is important to extend the conceptual framework from the $\Lambda =0$ case to the $\Lambda >0$ regime.

In the first paper in this series \cite{abk1} we began an exploration of this problem. Our findings related to gravitational waves can be summarized as follows. If one considers space-times which are asymptotically de Sitter in the sense introduced by Penrose \cite{rp} (more precisely, which satisfy Definition 2 in \cite{abk1}) then the asymptotic symmetry group is simply $\Diff$. Thus, with these boundary conditions, one cannot single out translations or rotations \emph{even asymptotically}.%
\footnote{As a result, in the quantum theory, we cannot decompose fields into positive and negative frequency parts even at $\scri$. Therefore, in contrast with the $\Lambda=0$ case, there are no candidate Hilbert spaces of asymptotic states which are necessary, e.g., to systematically discuss { whether the quantum evaporation of black holes is unitary.}}
Consequently, one cannot introduce 2-sphere charges analogous to the Bondi 4-momentum at $\scri$ \cite{bondi,sachs2}, or calculate fluxes of energy, momentum and angular momentum carried away by gravitational waves \cite{aams}. We then examined a common strategy to reduce $\Diff$ to the de Sitter group by strengthening the boundary conditions. The idea is to restrict oneself to those space-times for which the intrinsic 3-metric $q_{ab}$ on $\scri$ is conformally flat. This additional restriction seems natural, because the condition is satisfied in the familiar examples, including the Kerr-de Sitter and Friedmann-Lema\^itre space-times. Furthermore, the 2-sphere charges at $\scri$ associated with the Kerr-de Sitter time-translation and rotation yield the expected mass and angular momentum \cite{abk1}. 

However, we showed that the additional boundary condition is equivalent to demanding that \emph{the magnetic part $\B_{ab}$ of the leading order asymptotic Weyl curvature must vanish at $\scri$.} Now, in the case of Maxwell fields on asymptotically de Sitter space-times, the analogous requirement would be that the magnetic field $\B_{a}$ should vanish at $\scri$. This requirement would remove half the space of solutions \emph{by fiat}! {By analogy,} in the gravitational case, the strengthening of the boundary conditions appears to be physically unjustifiable. %@@ 
Furthermore, irrespective of whether one strengthens the boundary conditions in this manner or not, one does not have expressions of fluxes of energy-momentum and angular momentum carried away by gravitational waves. 
%
%\footnote{If one does require $\B_{ab}$ to vanish on $\scri$, then the 2-sphere charge integrals associated with the ten de Sitter asymptotic symmetries are well defined. But they are absolutely conserved. So, in these space-times gravitational waves would not carry any (de Sitter) energy, momentum or angular momentum \cite{abk1}!}
%
Indeed, for $\Lambda >0$, no gauge invariant characterization of gravitational waves is available in full general relativity! Thus, we have an apparent impasse: On the one hand the $\B_{ab}=0$ condition is too strong but, on the other hand, the boundary conditions are too weak without it (both for the gravitational radiation theory and quantum considerations).

A new framework is being constructed to overcome this difficulty and address related issues discussed in \cite{abk1}. In this paper we will complete the first step of that program by analyzing source-free, linearized gravitational waves in de Sitter space-time. In the $\Lambda=0$ case, the analogous analysis of linearized fields was necessary for the derivation of energy loss due to a time changing quadrupole moment in the weak field approximation (see, e.g., \cite{quadrupole}). More generally, it provided considerable intuition and important checks in the final construction of the theory of gravitational waves in exact general relativity \cite{bondi,rp,sachs1,sachs2,null-review}. In subsequent papers we will see that same is true in asymptotically de Sitter space-times. 

The main ideas of this paper can be summarized as follows. 

We will restrict ourselves to the (future) Poincar\'e patch of de Sitter space-time because, as we will see in \cite{ds4}, this provides the setting that is appropriate for describing isolated systems in full general relativity. The subgroup of the 10-dimensional de Sitter group that leaves this patch invariant is 7-dimensional, consisting of 4 (de Sitter) translations and 3 rotations; symmetries that enable one to define the total 4-momentum and angular momentum carried by test fields, including linearized gravitational waves. Because of the high degree of symmetry of the Poincar\'e patch, as is well-known, one can solve linearized Einstein's equation explicitly. By examining the behavior of solutions at $\scrip$ we will explicitly show that the (linearized analog of the) condition $\B_{ab}=0$ at $\scrip$ removes, by hand, half the number of degrees of freedom associated with gravitational waves. This will confirm the expectation from Maxwell's theory. 

For test matter, such as scalar, Maxwell or Yang-Mills fields, conserved quantities can be readily constructed using the stress-energy tensor. For linearized gravitational fields, on the other hand, we do not have a gauge invariant, local stress-energy tensor because in general relativity gravity is absorbed into space-time geometry. Therefore a new strategy is needed. A convenient route is provided by the covariant Hamiltonian framework where the phase space $\ps$ consists of solutions to linearized Einstein's equation. Diffeomorphisms generated by isometries have a well-defined action on $\ps$ which preserves the natural symplectic structure $\o$ on $\ps$. The Hamiltonians generating these canonical transformations provide us with formulas of energy-momentum and angular momentum carried by gravitational waves. We will express these quantities in terms of fields that are well-defined on $\scrip$. These expressions will be needed in the derivation of the energy loss due to a time-changing quadrupole moment in the $\Lambda >0$ case, derived in \cite{abk3}. 

Finally, we discuss the $\Lambda \to 0$ limit. Physically one expects that in this limit energy-momentum and angular momentum expressions should reduce to the well-known ones for linear gravitational waves in Minkowski space. However, the limit is delicate because of conceptually important discontinuities. In particular, while $\scrip$ is space-like for every $\Lambda >0$, it is null for $\Lambda =0$. Similarly, while the generator of every de Sitter `time translation' (used to define de Sitter energy) is space-like in a neighborhood of $\scrip$ for any $\Lambda >0$, it is time-like in a neighborhood of $\scrip$ for $\Lambda =0$. Consequently, while the flux of energy at de Sitter $\scrip$ can be \emph{arbitrarily} negative no matter how small $\Lambda$ is, it is strictly positive in the $\Lambda=0$ case. We provide a detailed, systematic procedure to take the limit and show that the de Sitter fluxes do go over to the Minkowski fluxes in the limit. This procedure will be useful { in reliably estimating the errors one makes by working in the asymptotically flat context rather than asymptotically de Sitter.}

The paper is organized as follows. In section \ref{s2} we collect results on the geometry of the Poincar\'e patch that will be used throughout our discussion and show that, for test Maxwell fields, the familiar fluxes of (the de Sitter) energy-momentum and angular momentum obtained using the stress-energy tensor can be derived using Hamiltonian methods that do not refer to the stress-energy tensor. In section \ref{s3} we study the asymptotic behavior of the explicit solutions to the linearized Einstein's equation in the Poincar\'e patch { and analyze the consequences of the $\B_{ab}=0$ condition.} In section \ref{s4} we introduce the covariant phase space $\ps$ of linearized gravitational fields in the Poincar\'e patch, derive expressions of Hamiltonians associated with the seven isometries, and express their limits to $\scrip$ using fields that have well-defined limits there. { In section \ref{s5}, we derive two properties of these fluxes. First, we show that when the subtleties associated with the $\Lambda \to 0$ limit are taken into account, our flux expressions of section \ref{s4} do reduce to the standard flux formulas associated with linear gravitational waves in Minkowski space-time. Second, we show that although gravitational waves in de Sitter space-time can carry arbitrarily large negative energy, for the class of solutions that are of direct physical interest in the investigation of isolated systems, they carry positive energy.}

Our conventions are as follows. Throughout we assume that the underlying space-time is 4-dimensional and the space-time metric has signature -,+,+,+.  The curvature tensors are defined via:  $2\nabla_{[a}\nabla_{b]} k_c = R_{abc}{}^d k_d$, $R_{ac} = R_{abc}{}^b$ and $R = R_{ab}g^{ab}$.

\section{Preliminaries}
\label{s2}

This section is divided into two parts: i) symmetries of the Poincar\'e patch; and, ii) the covariant phase space and conserved quantities associated with these symmetries.

\subsection{The Poincar\'e patch} 
\label{s2.1}

\begin{figure}[]
  \begin{center}
  \vskip-0.4cm
    %$a)$\hspace{8cm}$b)$
    \includegraphics[width=2.7in,height=2.7in,angle=0]{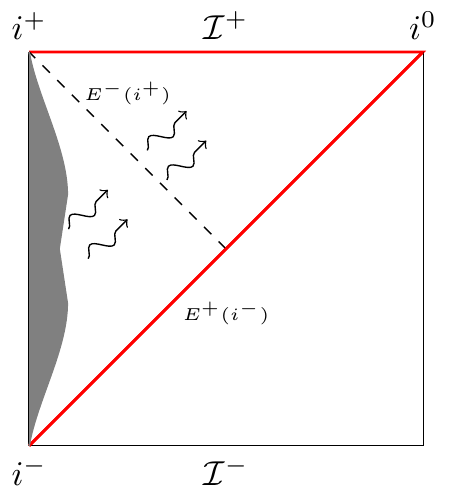}\hskip1.5cm
    \includegraphics[width=2.7in,height=2.7in,angle=0]{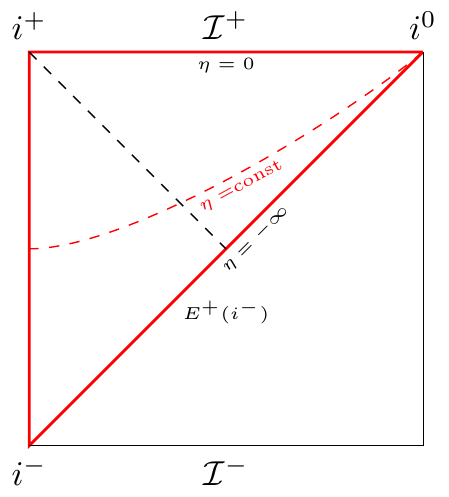}
\caption{\textit{Left Panel:} The Penrose diagram of a spherical isolated star in general relativity with $\Lambda>0$. The solid diagonal line denotes $E^{+}(i^{-})$, the future event horizon of $i^{-}$. The star and the radiation it emits are invisible to all observers whose world-lines are confined to the lower portion { of the de Sitter space-time below $E^{+}(i^{-})$. Therefore in the discussion of this isolated system, it is natural to restrict oneself to the upper half.}  The dashed diagonal line is $E^{-}(i^{+})$, the past event horizon of $i^{+}$. \textit{Right Panel:} The Poincar\'e patch of de Sitter space-time of interest is the upper triangle, to the future of the event horizon $E^{+}(i^{-})$ where $\eta=-\infty$. The $\eta= {\rm const}$ lines denote the cosmological slices, i.e., flat Cauchy surfaces.}
\label{poincare}
\end{center}
\end{figure}

In the $\Lambda=0$ case, to study isolated systems in the weak field limit, one investigates linearized gravitational fields in Minkowski space-time. For the $\Lambda >0$ case, it may seem natural to replace Minkowski space with de Sitter space-time. However, because of the differences in causal structures of these two space-times, an important difference arises. Consider an isolated system --such as a single star or a binary-- that is confined to a spatially bounded world-tube for all times (see the left panel in Fig. \ref{poincare}). In this case the matter world-tube has future and past end-points in both $\Lambda=0$ and $\Lambda >0$ cases, denoted by $i^{\pm}$. However, whereas in the $\Lambda=0$ case the future of $i^{-}$ is the entire Minkowski space-time, if $\Lambda >0$, it is only the future Poincar\'e patch of de Sitter. No observer whose world-line is confined to the past Poincar\'e patch can see the isolated system or detect the radiation it emits. Therefore, to study this system, it suffices to restrict oneself just to the future Poincar\'e patch rather than the full de Sitter space-time. Indeed, while it is difficult to impose the physically appropriate `no incoming radiation' boundary condition at $\scri^{-}$ \cite{bicak1}, as we will see in \cite{ds4}, %in full general relativity 
this condition can be naturally imposed at the cosmological horizon  $E^{+}(i^{-})$ that constitutes the past boundary of this Poincar\'e patch. Because our primary purpose is to develop intuition for the full, nonlinear theory, we will  restrict ourselves to this future Poincar\'e patch, although all our results can be readily extended to the full de Sitter space-time. 
%Left Panel: Figure needs $i^{+}$. E^{+}(i^{-}) should be a solid line. May be better to have a star that is radiating. So no KVFs and also no $H^{+}$. Only $H^{-}$ in bold red.
%Right Panel: Figure needs $i^{+}$. Remove the $\rho =const$ line and $\rho = -\infty$ label. Add a dashed line showing $E^{-}(i^{+})$ and KVFs
Next, for easy comparison with the rich literature on gravitational waves in cosmology, we will use coordinates $\eta, x,y,z$; with $x,y,z$ assuming their full range on $\mathbb{R}^{3}$ and the conformal time $\eta \in (-\infty,0)$ (see the right panel in Fig. \ref{poincare}). Then the de Sitter metric can be expressed as:
\be \label{desitter}
\bar{g}_{ab}\rmd x^{a}\rmd x^{b} \,=\, ({1}/{H\eta})^{2}\, \big(-\rmd \eta^{2} + \rmd \vec{x}{\,}^{2}\big) \, =:\, ({1}/{H\eta})^{2}\, \go_{ab}\rmd x^{a}\rmd x^{b}\, ,
\ee
where $H := \sqrt{\Lambda/3} =: 1/\ell$ is the Hubble parameter, the inverse of the cosmological radius $\ell$. 
\footnote{ These coordinates --as well as the coordinates $(t,\vx)$ discussed below-- have the disadvantage that they do not cover the past boundary of our Poincar\'e patch, i.e., the event horizon $E^{+}(i^{-})$ of $i^{-}$. But this limitation will not affect our considerations.}
{ While these coordinates are extremely convenient in the detailed calculations of gravitational perturbations, it is obvious that they are ill-suited for taking the limit $\Lambda \to 0$. To take this limit, it is simplest to use proper time $t$, which is related to the conformal time} $\eta$ via $H\eta = - e^{-Ht}$. In terms of $t$, the de Sitter metric becomes
\be \label{tx}
\bar{g}_{ab}\rmd x^{a}\rmd x^{b} \,=\, -\rmd t^{2} + e^{2Ht}\, \rmd \vec{x}{}\,^{2}\, 
\ee
and it is manifest that the metric coefficients go to those of Minkowski metric coefficients as $\Lambda$ goes to zero. Therefore, to compare geometric structures in de Sitter space-time to those in Minkowski space, it is important to use the differential structure induced on the Poincar\'e patch by $(t, \vec{x})$, and \emph{not} by $(\eta, \vec{x})$!

Locally, of course, this metric admits 10 (de Sitter) Killing fields. However, since the Poincar\'e patch is only a part of the de Sitter space-time, only those isometries are permissible that map this patch to itself. Therefore, we now have to restrict ourselves only to those  Killing fields that are tangential to its boundary $E^{+}(i^{-})$ in the full de Sitter space-time. As discussed in detail in section 4.C.2 of \cite{abk1}, these Killing fields constitute a 7-dimensional family. We have 3 spatial translations $T^{a}_{(i)}$ and 3 spatial rotations $R^{a}_{(i)}$, tangential to each $\eta = {\rm const}$ slice, generating the Euclidean group. In addition, there is a 7th Killing field
\be \label{T} T = - H\, \left[\eta \, \f{\partial}{\partial \eta} + x \f{\partial}{\partial x} + y \f{\partial}{\partial y} + z \f{\partial}{\partial z} \right]. \ee
We will refer to $T^{a}$ as \emph{time translation} because: i) it is the limit of the time translation Killing field in the Schwarzschild-de Sitter space-time as the mass goes to zero, and, ii) in the $(t, \vec{x})$ coordinates, it reduces to a time-translation in Minkowski space-time as $\Lambda \to 0$.% 
\footnote{Limit $\Lambda \to 0$ of $T^{a}$ illustrates the importance of using the correct differential structure to take this limit. Had we used the differential structure provided by $(\eta, \vec{x})$ we would have concluded from (\ref{T}) that $T^{a}$ vanishes in the limit. But this procedure would have been incorrect because the metric $\bar{g}_{ab}$ diverges in this limit (although it reduces to the well-defined Minkowski metric if the limit is taken using the differential structure induced by $(t, \vec{x})$). Note, incidentally, that $T^{a}$ is sometimes referred to as `dilation' because it is the conformal-Killing vector field representing a dilation with respect to the \emph{flat} metric $\go_{ab}$.}  
The commutation relations between these seven Killing fields are given by:
\be [T,T_{(i)}] =  H\, T_{(i)}\, , \quad  [T,R_{(i)}]= 0,\quad  [T_{(i)},R_{(j)}]
= \epsilon_{ij}{}^k\, T_{(k)}, \quad {\rm and} \quad [R_{(i)},R_{(j)}] = \epsilon_{ij}{}^k\, R_{(k)} \, . \label{7dimg}\ee
(Note that the time translation does \emph{not} commute with space-translations.) We will denote this 7-dimensional Lie-algebra of symmetries of the Poincar\'e patch by $\gp$ and the Lie group it generates by $\Gp$.%
\footnote{This is the group that leaves the point $i^{-}$ on $\scri^{-}$ of de Sitter space-time invariant. As $\Lambda \to 0$, $\Gp$ reduces to a well defined seven dimensional subgroup of the Poincar\'e group; the limit carries the memory of the preferred $t= {\rm const}$ slicing.} 
Finally, in the standard conformal completion of the Poincar\'e patch, $\scrip$ has $\mathbb{R}^{3}$ topology and this 7-dimensional group preserves the completeness of the allowed class of metrics on $\scrip$ \cite{abk1}.

\subsection{Maxwell fields in de Sitter space-time}
\label{s2.2}

As is well-known, each Killing symmetry $K^{a}$ leads to a conserved quantity. For matter fields --such as the Maxwell field $F_{ab}$-- the standard procedure is to use the stress-energy tensor $T_{ab} =\frac{1}{4 \pi} (F_{am}F_{bn}\, \bar{g}^{mn} \, - (1/4) \bar{g}_{ab} \, F_{cd} F_{mn}\, \bar{g}^{cm} \bar{g}^{dn}$) . The conserved quantity associated with a Killing field  $K^{a}$ is  given by
\be \label{stress}
\F_{K} = \int_{\M} T_{ab}\, K^{a} n^{b}\, \rmd^{3}V_{\M}  \ee
where the integral is taken over any Cauchy surface $\M$ with unit normal $n^{a}$. $\F_{K}$ may be regarded as the `flux' of the conserved quantity across $\M$.

However, for the linearized gravitational field, we do not have a gauge invariant, locally defined stress-energy tensor. We will now show that, in the Maxwell theory, the expression (\ref{stress}) of $\F_{K}$ can also be obtained using a covariant phase space framework without having to refer to the stress-energy tensor. In section \ref{s3}  we will use this alternate method to calculate conserved quantities for the linearized gravitational field.

Consider a globally hyperbolic space-time, $(\man, g_{ab})$ with a Killing field $K^{a}$. Denote by $\psM$ the space of all suitably regular, source-free solutions $F_{ab}$ to  Maxwell equations $\overline\nabla_{[a} F_{bc]} =0$ and {$\bar{g}^{ac} \overline\nabla_{c} F_{ab} =0$}. Starting from the Maxwell Lagrangian, one can show that $\psM$ is naturally endowed with a symplectic structure (i.e., a closed, non-degenerate 2-form) $\oM$:
\be \oM (F, \ub{F}) = {\frac{1}{4 \pi}} \int_{{\M}} {\bar{g}^{ac}} \big[{F}_{ab} \ub{A}{}_{c} - \ub{F}{}_{ab} {A}_{c}\,\big]\, n^{b}\, \rmd^{3}V_{\M}\, . \ee
Here $F$ and $\ub{F}$ are any two solutions to Maxwell equations,  $A_{a}$ is any vector potential for $F_{ab}$ (i.e., $F_{ab} = 2\,\nabla_{[a}A_{b]}$) and  $\M$ is again any Cauchy surface. Using Maxwell equations (and the suitable fall-off implicit in the regularity condition) it is easy to verify that the right side is independent of the choice of the Cauchy surface $\M$ and is gauge invariant. The pair $(\psM,\,\oM)$ is the Maxwell covariant phase space. Each Killing field $K^{a}$ on $\man$ naturally defines a vector field $\K$ on $\psM$ via: $\K\!\!\mid_{F} \, \equiv \delta_{K} F := \mathcal{L}_{K} F_{ab}$. Not surprisingly, the flow generated by $\K$ on $\psM$ preserves the symplectic structure $\oM$, i.e., defines a 1-parameter family of canonical transformations on $(\psM, \oM)$. The Hamiltonian generating this flow is a function $\H_{K}$ on $\psM$ given by:
\be \label{H1} \H_{K} := - \f{1}{2} \oM(F, \, \mathcal{L}_{K}F)\, . \ee
For any Killing field $K^{a}$ one can verify that $\H_{K}$ defined in (\ref{H1}) equals $\F_{K}$ defined in (\ref{stress}). (For details on the covariant phase space of fields, including general relativity, see, e.g., \cite{abr}.) 

Let us illustrate this result for the Killing fields in the Poincar\'e patch. Let us first set $K^{a} =S^{a}$, where $S^{a}$ stands for any one of the 6 Killing fields $T^{a}_{(i)}$ and $R^{a}_{(i)}$, tangential to the space-like slices $\M$ given by $\eta={\rm const}$.  Then, we have
\be \label{poynting} \F_{S} = {\frac{1}{4 \pi}} \int_{\M}\, (F_{am}F_{bn} \bar{g}^{mn} S^{a}n^{b})\, \rmd^{3}V_{\M} = {\frac{1}{4 \pi}} \int_{\M}\, (\epsilon_{abc} E^{b} B^{c} S^{a})\, \rmd^{3}V_{\M} \ee
where $E_{a} := F_{ab}n^{b}$ and $B_{a} := {}^{*}F_{ab}\, n^{b}$  are the electric and magnetic parts of the Maxwell field, and $\epsilon_{abc}$ the alternating tensor on the  slice $\M$. Thus, as one would expect, $\F_{S}$ is the flux of the $S$-component of the Poynting vector $\epsilon_{abc} E^{b}B^{c}$ across $\M$. Next, let us consider the Hamiltonian (\ref{H1}) generated by $S$:
\ba \H_{S} &=& {-\f{1}{8 \pi}} \int_{\M}\, {\bar{g}^{ac}} \big[(\mathcal{L}_{S} F_{ab}) A_{c}\, -\, F_{ab} (\mathcal{L}_{S} A_{c}) \big] n^{b}\, \rmd^{3}V_{\M}  \nonumber\\
&=& {-\f{1}{4 \pi}} \int_{\M}\, {\bar{g}^{bc}} F_{ac}(\mathcal{L}_{S}A_{b})\, n^{a}\, \rmd^{3}V_{\M} \, = \, {\f{1}{4 \pi}}
\int_{\M}\, (\epsilon_{abc} E^{b} B^{c} S^{a})\, \rmd^{3}V_{\M}\,  \ea
where in the first step we have integrated by parts and in the second step used Cartan identity and the Maxwell equation $\bar{D}_{a}E^{a} =0$. 
Thus, using the covariant phase space we can recover the conserved quantity $\F_{S}$ as the Hamiltonian $\H_{S}$ defined by the Killing symmetry $S^{a}$. Because of conformal invariance of Maxwell equations, we can easily take the limit as $\M$ approaches $\scrip$ and express the conserved flux as an integral over $\scrip$. The expression (\ref{poynting}) brings out the fact that if the magnetic field vanishes at $\scrip$, then that electromagnetic wave carries no angular momentum or linear momentum.

For the time translation $T^{a}$, the argument establishing the equality of $\F_{K}$ and $\H_{K}$ is the same but the calculation is a little more involved because $T^{a}$ has components both along \emph{and} orthogonal to the cosmological slices (see Eq.(\ref{T})). We find:
\be \F_{T} = \H_{T} = {\f{1}{8 \pi}} \int_{\M} \big[(E_{a}E_{b} + B_{a}B_{b})\, \bar{g}^{ab} + 2 \epsilon_{abc} E^{b} B^{c} T^{a}\big]\, \rmd^{3}V_{\M}.  \ee
In the limit as $\M$ approaches $\scrip$, $T^{a}$ becomes tangential to $\scrip$ (since $\eta =0$ at $\scrip$) and $\bar{g}^{ab}$ vanishes. Therefore the expression of the conserved energy reduces to an integral of the component of the Poynting vector along $T^{a}$:
\be \F_{T} = \H_{T} = {\f{1}{4 \pi}} \int_{\scrip} (\epsilon_{abc} E^{b} B^{c} T^{a})\, \rmd^{3}V_{\scrip}\,  \ee
where the electric and magnetic fields and the alternating tensor are calculated using any conformally rescaled metric that is regular at $\scrip$ (e.g., $\go_{ab}$). This expression {brings} out two interesting facts. First, in de Sitter space-time while the energy carried by electromagnetic waves is conserved as in Minkowski space-time, now it can be \emph{negative} and is \emph{unbounded below}. Second, if we restrict ourselves to Maxwell fields whose magnetic field  vanishes at $\scrip$, then those electromagnetic fields carry no energy either. Note that the second result is specific to $\scrip$: If the magnetic field vanishes on a cosmological slice $\eta= {\rm const} \not=0$, the energy of that Maxwell field does \emph{not} vanish unless the Maxwell field itself vanishes identically. The 3-momentum and the angular momentum, on the other hand do vanish. 

To summarize, for Maxwell fields, the conserved quantities associated with Killing fields in the Poincar\'e patch can be recovered as Hamiltonians on the covariant phase space, without any reference to the stress-energy tensor. Also, because all Killing fields $K^{a}$ on de Sitter space-time are tangential to $\scrip$ --and hence space-like--  one can express every conserved quantity $F_{K}$ as an integral across $\scrip$ of the component of the Poynting vector along $K^{a}$. This expression brings out the fact that if we were to require that  
the magnetic field vanish at $\scrip$, we would be left with electromagnetic waves \emph{that carry no 3-momentum or angular momentum, nor energy defined by de Sitter { isometries!}}\\ 

\section{Linearized gravitational fields}
\label{s3}

%This section is divided into two parts. In the first we discuss properties of solutions to the linearized Einstein's equation on de Sitter space-time and fix the terminology used in the rest of the paper. In the second, we introduce the covariant phase space $\ps$ of linearized gravitational fields.
%
%\subsection{Solutions and their asymptotic behavior}
% \label{s3.1}
As in section \ref{s2.1}, we will use the $(\eta,\, \vx)$ chart and the form (\ref{desitter}) of the de Sitter metric $\bar{g}_{ab}$ in the Poincar\'e patch. The perturbed metric will be denoted by $g_{ab}$,
\be \label{g1} g_{ab}\, =\, \bar{g}_{ab} + \epsilon\,\gamma_{ab}  \ee
where $\epsilon$ is the smallness parameter and $\gamma_{ab}$ denotes the first order perturbation. Then, in the Lorentz and radiation gauge, i.e., when the gauge freedom is exhausted by requiring that $\gamma_{ab}$ satisfy 
\be \label{gauge1} \overline{\nabla}{}^{a}\gamma_{ab} = 0;\quad \gamma_{ab}\eta^{a} =0;\quad {\rm and} \quad \gamma_{ab}\bar{g}^{ab} =0, \ee
the linearized Einstein's equation simplifies to
\be \label{lee1} \overline\Box\, \gamma_{ab} - 2 H^{2}\, \gamma_{ab}= 0 . \ee
(Here $\eta^{a}$ is a vector field normal to the cosmological slices with $\eta^{a}\partial_{a} = \partial/\partial\eta$.) Following a common strategy in the cosmology literature, it is convenient to rewrite (\ref{g1}) as
\be \label{g2}  g_{ab}\, \equiv \, a^{2}(\eta)\, (\go_{ab} +\epsilon\, h_{ab}) \,=\, \f{1}{(H\eta)^{2}}\,  (\go_{ab} +\epsilon\, h_{ab})\ee
since calculations are simpler in terms of the mathematical field $h_{ab}$ than in terms of the physical perturbation $\gamma_{ab}$. Indeed, the gauge conditions (\ref{gauge1}) can now be written using the background \emph{flat} geometry of $\go_{ab}$:
\be \label{gauge2} \nablao^{a} h_{ab} = 0; \quad h_{ab}\eta^{a} = 0; \quad {\rm and} \quad h_{ab}\go^{ab} = 0\, , \ee
and the linearized Einstein's equation becomes
\be \label{lee2} \boxo  h_{ab}- 2 \f{a^{\prime}}{a}\, h^{\prime}_{ab}  \,\,=\,\, \boxo\, h_{ab} + \f{2}{\eta}\, h^{\prime}_{ab} = 0\, , \ee
where $h_{ab}^{\prime} \equiv \eta^{c}\nablao_{c}h_{ab}$. Note that the gauge conditions and linearized Einstein's equation satisfied by $h_{ab}$ are the same as those satisfied by the linearized gravitational fields in Minkowski space-time in absence of a cosmological constant except for the extra term $(2/\eta)\, h^{\prime}_{ab}$ in the linearized Einstein's equation. In the $(t,\vec{x})$ differentiable structure that is well-suited to take the limit $\Lambda \to 0$, the extra term {$(2/\eta) \del_\eta h_{ab}= - 2 H \del_t h_{ab}$} goes to zero, just as one would expect.

As in the case of linearized fields in Minkowski space-time, it is simplest to find explicit solutions using a Fourier transform:
\be \label{ft} h_{ab}(\vx,\,\eta) \, \equiv\, \int \f{\rmd^{3}k}{(2\pi)^{3}}\, \sum_{(s)=1}^{2}\, \, h_{\vk}^{(s)} (\eta)\, \e^{(s)}_{ab}(\vk)\,\, e^{i\vk\cdot\vx}\ee
where $(s)$ labels the two helicity states and $\e^{(s)}_{ab}(\vk)$ are the polarization tensors, satisfying 
\ba \label{pol}
 \e^{(s)}_{[ab]}(\vk) &=& 0; \quad \e^{(s)}_{ab}(\vk)k^{b} = 0; \quad \e^{(s)}_{ab}(\vk)\qo^{ab} = 0;\nonumber\\
 (\e^{(s)}_{ab}(\vk))^{\star} &=& \e^{(s)}_{ab}(-\vk); \quad
 \quad \e^{(s)}_{ab}(\vk)\,\e^{(s^{\prime})}_{cd}(-\vk)\,\qo^{ac}\, \qo^{bd} = \delta_{(s),(s^{\prime})}\, . \ea
Here, and in what follows, $\qo_{ab}$ is the fixed spatial Euclidean metric on the cosmological slices, tailored to the co-moving coordinates $\vx$,\, and $\star$ denotes complex conjugation. The two functions $h_{\vk}^{(s)}(\eta)$  capture the gauge invariant information --the transverse traceless modes-- of the linearized gravitational field. Since $h_{ab}(\vx,\,\eta)$ are real fields, it follows that 
\be (h_{\vk}^{(s)}(\eta))^{\star} \, =\, h_{-\vk}^{(s)}(\eta)\, .  \ee
The field equation (\ref{lee2}) implies that the $h^{(s)}_{\vk}$ satisfy the ordinary differential equation (ODE):
\be \label{ode} (h_{\vk}^{(s)}){}^{\prime\prime} - \f{2}{\eta}(h_{\vk}^{(s)}){}^{\prime} + k^{2} h_{\vk}^{(s)}\, =\, 0, \ee
where the prime denotes differentiation with respect to $\eta$, and $k^{2} = \vk\cdot\vk$. The second order ODE (\ref{ode}) can be readily solved to obtain the general solution
\be \label{sol1} h_{\vk}^{(s)}(\eta) = (-2H)\,\big[ E^{(s)}_{\vk}\, (\eta\, \cos (k\eta) - (1/k)\,\sin (k\eta)) \, - \, B^{(s)}_{\vk} \, (\eta\sin (k\eta) + (1/k)\, \cos (k\eta)) \big] \ee
where $E^{(s)}_{\vk}$ and $B^{(s)}_{\vk}$ are arbitrary coefficients (in the Schwartz space), determined by the initial data of the solution. (These coefficients can also depend on $\Lambda$. We did not make this dependence explicit because in the main text we work with a fixed value of $\Lambda$.) Substituting (\ref{sol1}) in (\ref{ft}) we obtain the general solutions $h_{ab}$ representing first order perturbations. 

Next, let us discuss curvature. Since the Weyl tensor of de Sitter space-time vanishes, the first order perturbations ${}^{(1)}\!E_{ab}$ and ${}^{(1)}\!B_{ab}$ of the electric and magnetic parts of the Weyl curvature are gauge invariant and can be expressed directly in terms of the solutions $h_{\vk}(\eta)$ in (\ref{sol1}). To find these expressions, we first note that, in exact general relativity, the electric and magnetic parts are related to the first and second fundamental forms $q_{ab}$ and $K_{ab}$ on any space-like surface via
\ba \label{EB1} 
E_{ab} &=& \R_{ab} - K_{a}{}^{m}K_{mb} + K K_{ab} - (1/2) (q_{a}{}^{m}
q_{b}{}^{n} + q_{ab} q^{mn}) ({}^{4}\!R_{mn} - (1/6)\, {}^{4}\!R\, g_{mn})\nonumber\\
B_{ab} &=& \epsilon_{(a}{}^{mn}\, D_{|m|} K_{|n|b)}\ea
where $D$, $\epsilon_{abc}$ and $\R_{ab}$ are the derivative operator, alternating tensor and the Ricci curvature of the 3-metric $q_{ab}$, and ${}^{4}\!R_{ab}$ is the Ricci curvature of the space-time metric $g_{ab}$. 

It is straightforward to linearize these equations using the cosmological foliation on the de Sitter background. Calculations are simplified by noting that: (i) $E_{ab}$ and $B_{ab}$ are conformally invariant, and, (ii) a convenient conformal completion of de Sitter is provided by choosing the conformal factor $\Omega = -\, H\eta$, so that \emph{the conformal metric $\Omega^{2} \bar{g}_{ab}$ that is well behaved at $\scrip$ is just the Minkowski metric $\go_{ab}$ in the $(\eta,\,\vx)$ chart.} Therefore, in effect, linearization can be carried out using this flat background metric. The perturbed electric and magnetic parts of the Weyl tensor can be expressed using $h_{ab}$ and geometric structures associated with the flat 3-metric $\qo_{ab}$ on each cosmological slice:
\be {{}^{(1)}\!E_{ab} = - \f{1}{2}\, \big(\Do^{2} h_{ab} + \f{1}{\eta} \,h_{ab}^{\prime}\big)}, \quad {\rm and} \quad
{}^{(1)}\!B_{ab} = \f{1}{2}\mathring{\epsilon}_{(a}{}^{mn}\, \Do_{|m|} h^{\prime}_{|n|b)}. \ee
Recall that the boundary conditions at $\scrip$ imply that the Weyl curvature of an asymptotically de Sitter metric must vanish at $\scrip$ \cite{rp,abk1}. Therefore, the first order perturbations ${}^{(1)}\!E_{ab}$ and ${}^{(1)}\!B_{ab}$ of Weyl curvature also vanish at $\scrip$ and 
\be \Ecal_{ab} := \Omega^{-1}\,\,({}^{(1)}\!E_{ab}), \quad {\rm and} \quad 
\Bcal_{ab} := \Omega^{-1}\,\,({}^{(1)}\!B_{ab}) \ee
admit smooth limits there. We will refer to $\Ecal_{ab}$ and the $\Bcal_{ab}$ as the \emph{perturbed electric and magnetic parts of the Weyl curvature} as a short hand since it is these quantities that will feature in most of our discussion. Using explicit solutions (\ref{sol1}) it is easy to verify that they do indeed admit smooth limits to $\scrip$:
\ba \label{EB2}\Ecal_{ab} (\vx,\, \eta)\mid_{\eta=0}\, &=&\, \int \f{\rmd^{3}k}{(2\pi)^{3}}\, \sum_{(s)=1}^{2}\, \,k^{2}\, E_{\vk}^{(s)}\, \e^{(s)}_{ab}(\vk)\,\, e^{i\vk\cdot\vx}\nonumber\\
\Bcal_{ab} (\vx,\, \eta)\mid_{\eta=0}\, &=&\, \int \f{\rmd^{3}k}{(2\pi)^{3}}\, 
\sum_{(s)=1}^{2}\, \,k^{2}\, B_{\vk}^{(s)}\, {}^{*}\!\e^{(s)}_{ab}(\vk)\,\, e^{i\vk\cdot\vx}\ea
where ${}^{*}\!\e^{(s)}_{ab} = \epsilon_{a}{}^{mn}\, (k_{n}/k)\, 
\e^{(s)}_{mb}$ is the `dual' of the polarization tensor. These formulas bring out the meaning of the coefficients $E^{(s)}_{\vk}$ and $B^{(s)}_{\vk}$ that feature in the expression (\ref{sol1}) of a general solution to the linearized equations. $E_{\vk}^{(s)}$ directly determines the electric part of the perturbed Weyl tensor at $\scrip$ and $B_{\vk}^{(s)}$ the magnetic part at $\scrip$. 

It is therefore clear that the perturbed Weyl tensor has no magnetic part $\B_{ab}(\vec{x},\, \eta)$ \emph{at} $\scrip$ if and only if the solution $h_{ab}$ has the form
\be \label{sol2} h^{(s)}_{\vk}(\eta) = (-2H)\,\big[ E^{(s)}_{\vk}\, (\eta\, \cos k\eta - (1/k)\,\sin k\eta) \big] \ee 
everywhere, obtained by setting $B^{(s)}_{\vk} =0$ in (\ref{sol1}). Thus, the condition that the magnetic part vanish at $\scrip$ --or, that conformal flatness of the 3-metric at $\scrip$ be preserved to first order-- \emph{removes, by fiat, half the degrees of freedom from consideration.} The first expectation based on Maxwell fields is explicitly borne out. In section \ref{s4} we will show that the second expectation is also borne out: the remaining gravitational waves, that do preserve conformal flatness to first order, carry no energy, momentum or angular momentum.
\\

{\emph Remarks:}\\
(i) The explicit solution (\ref{sol1}) shows that, as one approaches $\scrip$ (i.e. as $\eta \to 0$), the term associated with $E^{(s)}_{\vk}$ vanishes while the term associated with $B^{(s)}_{\vk}$ survives. In the cosmology literature, the first is referred to as the `decaying mode' and the second as the `growing mode'. Thus, the requirement that the magnetic part of the perturbed Weyl curvature vanish at $\scrip$ removes by fiat the growing mode and leaves only the decaying mode. These perturbations $h_{ab}$ vanish at $\scrip$.

\noindent(ii) Let us return to the linearized Einstein's equation (\ref{lee2}) satisfied by $h_{ab}$. While one can think of $h_{ab}$ as a field propagating on the Minkowski metric $\go_{ab}$, because of the additional term $(2/\eta)\, h_{ab}^{\prime}$, the propagation is not sharp; there is a `tail term'. On the other hand, the linearized Weyl tensor satisfies conformally invariant equations. Its propagation does not have a tail term. Interestingly, the same is true of the time derivative of the metric perturbation: One can verify that it satisfies the conformally invariant equation, $(\overline{\Box} - ({}^{4}\!{\bar{R}}/6) ) h_{ab}^{\prime} =0$. Equivalently, since $\bar{g}_{ab} = (1/H^{2}\eta^{2})\, \go_{ab}$, it follows that $\boxo\, [(1/\eta)\, h_{ab}^{\prime}] =0$. Therefore it follows that the propagation of {$(1/\eta)\, h_{ab}^{\prime}$ on $(\man, \go_{ab})$, and hence of} $h^{\prime}_{ab}$ on $(\man,\, \bar{g}_{ab})$, is in fact sharp, without any tail terms. This fact has an interesting implication in the discussion of the quadrupole formula \cite{abk3}.

\section{The Hamiltonian framework} 
\label{s4}

This section is divided into two parts. In the first, we construct the covariant phase space of source-free, linearized gravitational fields on the de Sitter background. In the second, we obtain expressions of energy, momentum and angular momentum carried by gravitational waves by computing the Hamiltonians corresponding to the seven Killing fields on the Poincar\'e patch.

\subsection{The covariant phase space}
\label{s4.1}   

For linearized gravitational fields, the covariant phase space $\ps$ can be taken to be the space of solutions $\gamma_{ab}$ to the equations (\ref{gauge1}) and (\ref{lee1}). For simplicity, we will assume that the solutions of interest have initial data { in the Schwartz space of rapidly decreasing, smooth fields,} although these conditions can be weakened considerably. The standard procedure (see, e.g. \cite{abr}) endows $\ps$ with a symplectic structure $\omega$. 
%
%\label{sym}
%\be \omega(h,\, \ub{h})\, =\, \f{1}{2\kappa}\, \int_{\M} \rmd^{3}V\, \big( h_{ab}\, \ub{p}_{cd} - \ub{h}_{ab} p_{cd}\big) q^{ac} q^{bd}   \ee
%
%where $q_{ab}$ is the 3-metric on $\M$ induced by the de Sitter metric $\bar{g}_{ab}$, $p_{ab}$ is the linearization of the `momentum field' $\sqrt{q}\, (K_{ab} - K q_{ab})$..
%
Restricted to the cosmological slices $\M$ (given by $\eta = {\rm const})$, it becomes:
\be \label{sym1} \omega(\gamma,\ub{\gamma}) \, \equiv\, \omega(h,\, \ub{h})\, :=\, \f{a^{2}(\eta)}{{4}\kappa}\, \int_{\M} \rmd^{3}x\, \big(h_{ab}\, \ub{h}^{\prime}_{cd}\,\, -\,\,  {h}_{ab}^{\prime}\, \ub{h}_{cd}\big)\,\, \qo^{ac}\, \qo^{bd}\, , \ee
where $h_{ab}$ is related to the physical metric perturbation $\gamma_{ab}$ via $\gamma_{ab} = a^{2}\, h_{ab}$ (see Eq. (\ref{g2})) {and $\kappa = 8 \pi G$}. It is easy to verify that (\ref{gauge2}) and (\ref{lee2}) imply that the integral is independent of the $\eta={\rm const}$ slice on which it is evaluated. This form of the symplectic structure is useful in calculations within the Poincar\'e patch. Furthermore, as we will see in section \ref{s4.2}, it is well-adapted for taking the limit $\Lambda \to 0$.

In the cosmology literature, one often works with the functions $h_{\vk}^{(s)}(\eta)$ defined in (\ref{ft}) and their Fourier transforms
\be \phi^{(s)} (\vx, \eta)\, := \, \f{1}{\sqrt{4\kappa}}\, \int \f{\rmd^{3}k}{(2\pi)^{3}}\,  \, h_{\vk}^{(s)} (\eta)\, e^{i\vk\cdot\vx}\ee
in place of the tensor fields $\gamma_{ab}$ or $h_{ab}$. {(The factor of $\sqrt{4\kappa}$  is introduced to endow $\phi^{(s)}$ with the standard dimensions of a scalar field, so that the scalar and tensor perturbations can be treated in a completely parallel manner. See, e.g., section 3.D of \cite{aan2}.)} These are referred to as the two \emph{tensor modes.} It is straightforward to verify that these fields satisfy the wave equation in de Sitter space-time
\be\label{wave} \overline{\Box}\, \phi^{(s)}(\vx, \eta) \, =\, 0\, .  \ee
Thus, each tensor mode $\phi^{(s)}$ of the linearized gravitational field satisfies just the massless Klein-Gordon equation. It is clear that, given fixed polarization tensors $\e^{(s)}_{ab} (\vk)$, there is a natural isomorphism between the functions $\phi^{(s)}(\vx, \eta)$ and solutions $h_{ab}(\vx,\eta)$ to the linearized Einstein equation (\ref{lee2}) and gauge conditions (\ref{gauge2}). It is easy to check that the symplectic structure (\ref{sym1}) on $\ps$ translates to the standard symplectic structure on the covariant phase space $\ps^{\rm KG}$ consisting of pairs of solutions $\phi \equiv \{\phi^{(s)}(\vx,\eta)\}$ to the Klein-Gordon equation: 
\be \label{sym3} \omega_{\rm KG}(\phi, \ub\phi) = {a^{2}(\eta)} \int_{\M} \rmd^{3}x\, \sum_{(s)=1}^{2}\,\big(\phi^{(s)} \,(\ub{\phi}^{(s)})^{\prime}\, -\, (\phi^{(s)})^{\prime}\, \ub{\phi}^{(s)}\big)\, . \ee
This form of the symplectic structure is useful to compute expressions of fluxes of energy-momentum and angular momentum that are adapted to the `tensor modes' used in the cosmological perturbation theory. %The disadvantage is that, as with the expression (\ref{sym1}) of the symplectic structure, the limit to $\scrip$ is now difficult because the pre-factor $a^{2}(\eta)$ diverges there. 

However, expressions (\ref{sym1}) and (\ref{sym3}) of the symplectic structure have one drawback: because of the multiplicative factor $a^{2}(\eta) = (1/H^{2}\eta^{2})$, they are not well-suited to take the limit to $\scrip$ (where $\eta=0$). While, the limit itself is well defined because the symplectic structure is independent of $\eta$, to express physical results --e.g. the formula of energy-- in terms of fields that are well defined at $\scrip$, one has to be extremely careful in keeping track of terms in the integrand which tend to zero at the appropriate rate to compensate for the apparent blow up as $1/\eta^{2}$ due to the pre-factor in front of the integral. Also, these expressions are not gauge invariant as they use specific gauge conditions (\ref{gauge2}). To overcome these limitations, it is convenient to recast the expression (\ref{sym1}) using the relation between the perturbed electric part of the Weyl tensor $\Ecal_{ab}$ and the metric perturbation,
\be \label{E} 2\, \Ecal_{ab}(\vx,\,\eta) = \f{1}{H\eta^{2}}\, \big(h^{\prime}_{ab} {+}  \eta\,\Do^{2} h_{ab}\big)\, ,  \ee
that holds on any cosmological slice. Substituting for $h_{ab}^{\prime}$ in terms of $\Ecal_{ab}$ and simplifying by performing integrations by parts, we obtain: 
\be \label{sym2} \omega(h,\, \ub{h})\, =\, \f{1}{{2}H\kappa}\, \int_{\M} 
\rmd^{3}x\, \big( h_{ab}\, \underline{\Ecal}_{cd}\, - \,\ub{h}_{ab}\, \Ecal_{cd}\big)\,\, \qo^{ac}\, \qo^{bd}\, .\ee
We will use both expressions, (\ref{sym1}) and (\ref{sym2}), of the symplectic structure  on $\ps$ in our discussion of the conserved fluxes associated with the 7 Killing vectors. (The equivalent form (\ref{sym3}) in terms of the Klein-Gordon fields $\phi^{(s)}$ turns out not to be as useful in providing hints for the full, nonlinear theory.) 

We will conclude this discussion by pointing out several consequences that follow immediately from the form (\ref{sym2}) of the symplectic structure. First, it is transparent that $(1/{2} H\kappa)\,\Ecal^{ab}$ can be regarded as the momentum that is canonically conjugate to the metric perturbation $h_{ab}$. Second, as we saw in section \ref{s3}, the perturbations $h_{ab}$ as well as the perturbed electric part of the Weyl tensor $\Ecal_{ab}$ admit well defined limits to $\scrip$. Therefore, one can take the limit $\M \to \scrip$ simply by evaluating the integral (\ref{sym2}) on $\scrip$. This feature will facilitate our task  of expressing energy, momentum and angular momentum in terms of asymptotic fields at $\scrip$. In turn, these expressions will be directly useful in \cite{abk3} to obtain a formula for the energy emitted by a time changing quadrupole, and establishing its positivity. The third and more important feature is gauge invariance. Note first that $\Ecal_{ab}$ by itself is gauge invariant, it is tangential to the cosmological slices, and it is divergence-and trace-free. This fact enables us to drop the gauge fixing conditions (\ref{gauge2}) and consider general perturbations. For, if either $\gamma_{ab}$ (or, $\ub{\gamma}{}_{ab}$) is a pure gauge field  --i.e. of the form $\bar\nabla_{(a} \xi_{b)}$ for a space-time vector field $\xi^{a}$-- properties of $\Ecal_{ab}$ ensure that the expression (\ref{sym2}) of $\omega(h,\, \ub{h})$ vanishes identically. Thus, the passage from $h_{ab}^{\prime}$ to $\Ecal_{ab}$ using (\ref{E}) has provided us with a manifestly gauge invariant expression (\ref{sym2}) of the symplectic structure. Finally, using the explicit solutions (\ref{sol1}), we can re-express the symplectic structure in terms of the coefficients $E_{\vk}$ and $B_{\vk}$:
\be \label{sym4} 
\omega(h,\, \ub{h}) = \f{{1}}{\kappa}\, \int \f{\rmd^{3}k}{(2\pi)^{3}}\, \sum_{(s)=1}^{2}\, \,k\, \big((B^{(s)}_{\vk})^{\star}\, \ub{E}^{(s)}_{\vk} \, -\, (\ub{B}_{\vk}^{(s)})^{\star}\, E_{\vk}^{(s)}\big)\, , \ee
where, as before, $\star$ denotes complex-conjugation. Consequently, the pull-back of the symplectic structure to the subspace of $\ps$ on which $\Bcal_{ab}$ vanishes on $\scrip$ --or, alternatively, on which $\Ecal_{ab}$ vanishes on $\scrip$-- is  identically zero. These subspaces are among the maximal Lagrangian subspaces of $\ps$. In this respect the situation is again completely parallel to that in the Maxwell theory.

\subsection{3-momentum, angular momentum and energy carried by gravitational waves}
\label{s4.2}

We can now calculate the Hamiltonians on $\ps$ corresponding to the seven Killing fields on the Poincar\'e patch. Recall from (\ref{g2}) that the physical metric perturbation is $\gamma_{ab} = a^{2} h_{ab}$ and it satisfies the gauge conditions (\ref{gauge1}) and linearized Einstein's equation (\ref{lee1}) that refer only to the background de Sitter metric $\bar{g}_{ab}$. Therefore, if $\gamma_{ab}\in \ps$, then so is ${\gamma}^{(K)}_{ab} :=\mathcal{L}_{K}\, \gamma_{ab}$, for any  Killing field $K^{a}$ of $\bar{g}_{ab}$. From the definition (\ref{g2}) of $h_{ab}$, it follows that  
\be \gamma^{(K)}_{ab} = a^{2}\, (\mathcal{L}_{K}\, h_{ab} + 2(a^{-1}\,\mathcal{L}_{K} a)\,h_{ab}) =: a^{2} h^{(K)}_{ab}\ee
with $a = -1/(H\eta)$. As in the Maxwell case, the isometries generated by each of the seven Killing fields $K^{a}$ in $\gp$ provide a 1-parameter family of canonical transformations on $\ps$. From general results on the covariant phase space \cite{abr} it follows that the corresponding Hamiltonian is again given by
\be \label{H2} \H_{K} : = - \f{1}{2}\, \omega(\gamma,\, \gamma^{(K)}) = - \f{1}{2}\, \omega(h,\, h^{(K)}).\ee
Recall from section \ref{s3} that \emph{if $\Bcal_{ab} =0$ at $\scrip$, then $h_{ab}$ also vanishes there.} In this case, then, we have $\H_{S} =0$. Thus, although there do exist linearized  gravitational waves that retain conformal flatness of the induced geometry at $\scrip$ to first order, \emph{they carry no energy, 3-momentum or angular momentum.}

We will now compute the Hamiltonians (\ref{H2}) for the seven Killing fields in $\gp$.

\subsubsection{3-momentum and angular momentum}
\label{s4.2.1}

As in the case of Maxwell fields discussed in section \ref{s2.2}, the calculations are identical for the 3 spatial translations $T^{a}_{(i)}$ and the 3 rotations $R^{a}_{(i)}$. Let us therefore again denote by $S^{a}$ any of these six Killing fields and calculate the 3-momentum or angular momentum $\H_{S}$, and then discuss energy $\H_{T}$ separately. For these six Killing fields, we have $h^{(S)}_{ab} = \mathcal{L}_{S}\, h_{ab}$ since these fields are all tangential to the $\eta={\rm const}$ surfaces. Furthermore, from (\ref{E}) it follows that the corresponding perturbed electric part of the Weyl tensor, $\Ecal^{(S)}_{ab}$, is given by: 
\be \Ecal_{ab}^{(S)} = \f{1}{2H\eta^{2}}\, \big(\mathcal{L}_{\eta} (\mathcal{L}_{S} h_{ab})\, {+} \,\eta\,\Do^{2} (\mathcal{L}_{S} h_{ab})\, \big)\,=\, \mathcal{L}_{S}\, \Ecal_{ab} \ee
Therefore (\ref{H2}) becomes:
\ba \label{H3} \H_{S} &=&  - \f{1}{2}\, \omega(h,\, h^{(S)}) = -\f{1}{{4}H\kappa}\,\int_{\M} \rmd^{3}x \, \big(h_{ab}\,\Ecal_{cd}^{(S)}\, -\, h_{ab}^{(S)}\,\Ecal_{cd}\big)\, \qo^{ac}\, \qo^{cd}\nonumber\\
&=& \f{1}{{2}H\kappa}\,\int_{\M} \rmd^{3}x \, \big(\Ecal_{ab}\,\, \mathcal{L}_{S}h_{cd}\big)\, \qo^{ac}\, \qo^{cd}\, , \ea
where, in the second step we have integrated by parts. Thus, the expressions of 3-momentum and angular momentum mirror those in the Maxwell theory. Since the integrand in (\ref{H3}) refers only to the fields $h_{ab}$,\, $\Ecal_{ab}$ and the metric $\qo_{ab}$, all of which have smooth limits to $\scrip$, to take the limit $\M \to \scrip$ we just have to evaluate (\ref{H3}) on $\scrip$. 

Finally, let us consider the limit $\Lambda \to 0$ of $\H_{S}$. Since the Hubble parameter $H$ tends to zero in this limit, from the form of (\ref{H3}), the limit seems divergent at first sight. However, this conclusion is incorrect because fields in the integrand also depend on $H$. Let us therefore analyze the limit more carefully. As explained in section \ref{s2}, to take this limit, we should use the differential structure induced by the chart $(t,\, \vx)$ on the Poincar\'e patch (and not by the chart $(\eta,\, \vx)$). Then, for the background geometry, we find that as $\Lambda \to 0$, we have
\ba \label{lim1} \bar{g}_{ab} &\to & \eta_{ab} = - \partial_{a} t \,\partial_{b} t + \partial_{a} x\, \partial_{b} x + \partial_{a} y\, \partial_{b} y +\partial_{a} z\, \partial_{b} z;\nonumber\\ 
 H\eta \equiv - e^{-Ht} &\to&  -1;\quad\quad\,\,\, \f{\partial}{\partial\eta} \to    \f{\partial}{\partial t} \equiv t^{a} \partial_{a};\quad \quad\,\,\,    T^{a} \to  t^{a}\, . \ea
Note that the Minkowski metric $\eta_{ab}$ in (\ref{lim1}) is distinct from the Minkowski metric $\go_{ab}$ in (\ref{g2}). In the Poincar\'e patch, each of the Cartesian coordinates,\, $(t,\, \vx)$\, of $\eta_{ab}$ takes the full range of values, $(-\infty, \, \infty)$ (whereas $\eta$, of $\go_{ab}$ only runs from $(-\infty, \, 0)$). A second important point for the limit is that the $(t,\, \vx)$ chart does not cover $\scrip$ (where $t=\infty$). Therefore, to take the $\Lambda \to 0$ limit, we are led to evaluate the symplectic structure and Hamiltonians $\H_{S}$ on a cosmological slice corresponding to a finite constant value of $t$. 

Consider any 1-parameter family of smooth fields $h_{ab}(\Lambda)$ which solve the gauge condition (\ref{gauge2}) and the linearized Einstein equation (\ref{lee2}) for each $\Lambda$ and admit a smooth limit $\ho_{ab}$ as $\Lambda \to 0$ everywhere on the Poincar\'e patch.
Then, using (\ref{lim1}), it is straightforward to show that $\ho_{ab}$  
satisfies the Lorentz and radiation gauge, as well as the linearized Einstein's equation with respect to the Minkowski metric $\eta_{ab}$:
\be \label{hoeqns} \partial^{a} \ho_{ab} =0;\,\, \ho_{ab}\eta^{ab} =0; \,\, \ho_{ab} t^{b} =0;\,\,\, {\rm and}\,\,\,\, \partial^{c}\partial_{c}\, \ho_{ab} =0\, . \ee
Denote the space of solutions $\ho_{ab}$ to these equations by $\pso$.
It is straightforward to verify that in the limit $\Lambda \to 0$, the symplectic structure $\omega$ on $\ps$ goes over to the standard symplectic structure $\omegao$ on $\pso$:
\be \omegao(\ho,\, \ub{\ho}) = \f{1}{4\kappa}\, \int_{\M} \rmd^{3}x\,\big(\ho_{ab}\, \partial_{t}(\ub{\ho}_{cd})\, -\, \partial_{t}(\ho_{ab})\, \ub{\ho}_{cd}\big)\qo^{ac}\, \qo^{bd} \, ,\ee
where the integral is taken on a $t= {\rm const}$ slice. Finally, the limit $\Ho_{S}$ of the Hamiltonian $\H_{S} = (-1/2)\,\,\omega (h,\, h^{(S)})$ is given by:
\ba \Ho_{S} &=&  -\f{1}{2}\, \omegao(\ho,\, \mathcal{L}_{S}\ho )\nonumber\\
&=& \f{1}{4\kappa}\, \int_{\M} \rmd^{3}{x}\, \big(\partial_{t}(\ho_{ab})\,\,\, \mathcal{L}_{S}\ho_{cd}\big)\, \qo^{ac}\, \qo^{bd}\, . \ea 
This is precisely the expression of the linear and angular momentum of linearized gravitational waves in Minkowski space-time. Thus, although the procedure of taking the limit $\Lambda \to 0$ is rather subtle, the de Sitter 3-momentum and angular momentum (\ref{H3}) do reduce to the standard conserved fluxes in Minkowski space-time.\\

\emph{Remark:}\\
While taking the limit, we assumed the existence of a family $h_{ab}(\Lambda)$, satisfying the gauge conditions (\ref{gauge2}) and linearized Einstein equation (\ref{lee2}) for each $\Lambda$, that admits a smooth limit $\ho_{ab}$ on the Poincar\'e patch as $\Lambda \to 0$. An explicit example of such a family is provided by setting in (\ref{sol1}) 
\ba E^{(s)}_{\vk}(\Lambda) &=& C^{(s)}_{\vk} \, \sin(k/H)\, +\, D^{(s)}_{\vk}\, \cos(k/H);\,\, \nonumber\\
 B^{(s)}_{\vk}(\Lambda) &=&  -C^{(s)}_{\vk} \, \cos(k/H)\, +\, D^{(s)}_{\vk}\, \sin(k/H)\, . \ea
A careful calculation shows that in the limit $\Lambda \to 0$ the field $h^{(s)}_{\vk}(\Lambda)$ of (\ref{ft}) goes over to 
\be \ho^{(s)}_{\vk} = 2 C^{(s)}_{\vk}\, \sin kt + 2 D^{(s)}(\vk)\, \cos kt \ee
and the limit $\ho_{ab}$ of the family $h_{ab}(\Lambda)$ is given by
\be \label{mink-limit} \ho_{ab} (\vx, t) \, =\, \int \f{\rmd^{3}k}{(2\pi)^{3}}\, \sum_{(s)=1}^{2}\, \big( A^{(s)}_{\vk}\, e^{ikt}\, +\, (A^{(s)})^{\star}(\vk)\, e^{-ikt} \big)\, \e^{(s)}_{ab}(\vk)\,\,  e^{i{\vk}\cdot{\vx}}\ee  
where $2A^{(s)}_{\vk} = D^{(s)}_{\vk} - i\, C_{\vk}$. Thus, a general solution to the linearized Einstein's equation in Minkowski space in the transverse, traceless, radiation gauge can be obtained as a limit of this family $h_{ab}(\Lambda)$.

\subsubsection{Energy}
\label{s4.2.2}

Next, let us consider the energy $\H_{T}$ defined by the time translation $T^{a}$ of (\ref{T}). In this case, the calculation is not as straightforward because: (i) the vector field $T^{a}$ is \emph{not} tangential to the cosmological slices $\eta={\rm const}$ except at $\scrip$; (ii) $h_{ab}^{(T)}$ has an extra term relative to $h_{ab}^{(S)}$,
\be h_{ab}^{(T)} = \mathcal{L}_{T}\, h_{ab}\, +\, {2H}\, h_{ab}\, ; \ee
and, (iii) a detailed calculation shows that $\Ecal^{(T)}_{ab}$ also has an extra term:
\be \Ecal^{(T)}_{ab} = \mathcal{L}_{T}\, \Ecal_{ab} - H \Ecal_{ab}\, . \ee 
Once these differences are taken into account, the conserved energy-flux $\H_{T}$ across $\M$ can be calculated using (\ref{H2}). We have:
\be \label{H4} \H_{T} = -\f{1}{4H\kappa}\,\int_\M \rmd^{3}x \, \big(h_{ab} \mathcal{L}_{T} \Ecal_{cd} - \Ecal_{cd} \mathcal{L}_{T} h_{ab} -{3} H h_{ab}\Ecal_{cd}\big)\, \qo^{ac}\qo^{bd}\,  \ee
where, as before, $\M$ is any cosmological slice. However, since $T$ is not tangential to $\M$, on a general cosmological slice we cannot integrate by parts as we did for $\H_{S}$. Again the limit to $\scrip$ is straightforward since all fields in the integrand have smooth limits to $\scrip$. Furthermore, in the limit $T^{a}$ becomes tangential to $\scrip$ enabling us to simplify (\ref{H4}) further:
\ba \label{H5} \H_{T} &=& \f{1}{2H\kappa}\,\int_{\scrip}\, \rmd^{3}x\, \Ecal_{cd}\,\big(\mathcal{L}_{T} h_{ab} +2H\,h_{ab}\big)\, \qo^{ac}\qo^{bd}\, \\
&=& \f{{H}}{\kappa} \int \f{\rmd^{3}k}{(2\pi)^{3}}\, k \,\,{\sum_{(s)=1}^{2}} \left(E^{(s)}_{\vk} \mathcal{L}_k (B^{(s)}_{\vk})^{\star} + 2\, E^{(s)}_{\vk} \, (B^{(s)}_{\vk})^{\star} \right)\, , \ea
where in the second step we have used the explicit solutions (\ref{sol1}) for $h_{ab}$ in terms of Fourier modes. { Since $\Bcal_{ab}=0$ if and only if $B^{(s)}_{\vk} =0$, the last expression makes it explicit that if a gravitational wave does not change conformal flatness of the intrinsic geometry at $\scrip$ to first order, it does not carry  energy.} Finally, the expression (\ref{H2}) of $\H_{K}$ is linear in $K^{a}$ for all Killing fields. Therefore, $\H_{\lambda T} = \lambda \H_{T}$ for all real numbers $\lambda$. For linearized gravitational waves on Minkowski space-time, energy is positive definite and vanishes if and only if the perturbation is pure gauge. On the de Sitter space-time, the conserved energy $\H_{T}$ can have either sign and we have an infinite dimensional subspace of the physical, transverse-traceless modes for which the energy vanishes. From (\ref{H5}) it is clear that energy also vanishes if $\Ecal_{ab}$ vanishes on $\scrip$. (The other possibility, $\mathcal{L}_{T} h_{ab} = -2H h_{ab}$ on $\scrip$ is not realized because such perturbations would not { be in the Schwartz space on} $\scrip$ of the Poincar\'e patch, which is topologically $\mathbb{R}^{3}$ \cite{abk1}.)

Finally, let us consider the limit $\Lambda \to 0$ of the conserved energy $\H_{T}$. For reasons given in section \ref{s4.2.1}, we have to use the differential structure induced by the coordinates $(t,\, \vx)$ and work with a cosmological slice in the Poincar\'e patch with $\eta\not=0$. Let us again suppose that we have a 1-parameter family of perturbations $h_{ab}(\Lambda)$ that satisfy the gauge conditions (\ref{gauge2}) and the linearized Einstein equation (\ref{lee2}), and admit a smooth limit $\ho_{ab}$ as $\Lambda \to 0$. As discussed above, $\ho_{ab}$ is a metric perturbation on the Minkowski metric $\eta_{ab}$, satisfying (\ref{hoeqns}).  The limit $\Ho_{t}$ of the Hamiltonian $\H_{T} = (-1/2)\,\,\omega (h,\, h^{(T)})$ is given by:
\ba \Ho_{T} &=& - \f{1}{{8}\kappa}\, \int \rmd^{3}x\, \big(\ho_{ab} \partial_{t}^{2} \ho_{cd}\,  - \, (\partial_{t}\ho_{ab})(\, \partial_{t}\ho_{cd})\big)\, \qo^{ac}\qo^{bd}\nonumber\\
&=& \f{1}{{8}\kappa}\, \int \rmd^{3}x\, \big((\partial_{t}\ho_{ab})\, (\partial_{t}\ho_{cd}) + (\Do_{m}\ho_{ab})\, (\Do_{n}\ho_{cd})\, \qo^{mn}\big)\, \qo^{ac}\qo^{bd}\, . \ea
where in the second step we have used (\ref{hoeqns}) and integrated by parts. This is precisely the conserved energy flux of the linearized gravitational field $\ho_{ab}$ in Minkowski space-time. Thus, our energy expression (\ref{H4}) for linearized gravitational fields in de Sitter space-time does have the expected limit as $\Lambda \to 0$. Note that the limit is quite subtle and discontinuous: \emph{While $\H_{T}$ can be negative and arbitrarily large, no matter how small the positive $\Lambda$ is, in the limit $\Lambda \to 0$ we obtain $\Ho_{T}$ which is positive definite!} Geometrically, this occurs because while the Killing field $T^{a}$ of de Sitter metric $\bar{g}_{ab}$ is space-like in the `upper half of the Poincar\'e patch' for every $\Lambda >0$, its limit, the Killing field $t^{a}$ of $\eta_{ab}$, is time-like everywhere.\\

\emph{Remarks}:\\
(i) In the cosmological literature, the discussion of `energy' often refers to the Hamiltonians $\H_{\eta}$ or $\H_{t}$ that generate evolution along the conformal time $\eta$ or proper time $t$. Since $\eta^{a}$ and $t^{a}$ are not Killing fields, these Hamiltonians are \emph{not} conserved. Thus, they are unrelated to the conserved energy $\H_{T}$ discussed above and are \emph{not} the analogs of the standard notion of energy in Minkowski space-time, used in the gravitational radiation theory.

\noindent(ii) As discussed in section \ref{s4.1}, in cosmology one often encodes the metric perturbations $\gamma_{ab}(\vx,\eta)$ in the two `tensor modes' $\phi^{(s)}(\vx,\eta)$ that satisfy the Klein-Gordon equation with respect to the de Sitter metric $\bar{g}_{ab}$. On the Klein-Gordon phase space $\ps^{\rm KG}$, the isometry generated by any Killing field $K^{a}$ again defines a 1-parameter family of transformations that preserve the symplectic structure $\omega_{\rm KG}$. As one would expect, the corresponding Hamiltonians agree with the $\H_{K}$ obtained above for all seven Killing fields. That is, our energy-momentum and angular momentum expressions $\H_{T}$ and $\H_{S}$ hold both for the metric perturbations $\gamma_{ab}$ satisfying (\ref{gauge1}) and (\ref{lee1}) and the `tensor modes' $\phi^{(s)}$ satisfying the wave equation (\ref{wave}) in the Poincar\'e patch. 

\noindent(iii) Finally, we note that the explicit solutions (\ref{sol1}) are widely used in the cosmological literature on linearized gravitational waves.  However, the primary interest there is on the effect of these gravitational waves on the polarization of the CMB electromagnetic waves. To our knowledge, this literature does not contain the analysis of the asymptotic behavior of these perturbations at $\scrip$, or the implications of the assumption that the perturbations preserve conformal flatness of $\scrip$ to linear order, nor a discussion on the isometry group $\Gp$ that preserves the Poincar\'e patch, or the associated conserved fluxes $\H_{K}$ given above.

\section{Discussion}
\label{s5}

In the $\Lambda =0$ case, there is a well-developed theory of isolated systems and gravitational radiation in full, nonlinear general relativity that has played a dominant role in a number of areas of gravitational science. In the first paper \cite{abk1} in this series, we showed that there are significant conceptual obstacles in extending this theory to allow a positive cosmological constant, however small, because the limit $\Lambda \to 0$ is discontinuous. In particular, whereas $\scri$ is space-like, no matter how small $\Lambda$ is, it is null when $\Lambda$ vanishes. If $\Lambda$ were zero and the accelerated expansion of the universe is caused by some matter field rather than a cosmological constant, that field would not have the asymptotic fall-off we are familiar with in the $\Lambda=0$ case, and space-time curvature far away from the sources would be similar to that in asymptotically de Sitter space-times. Therefore, difficulties discussed in \cite{abk1} would persist also in the $\Lambda=0$ case { if the observed accelerated expansion continues to infinite future.} To overcome these obstacles, one needs a new framework. In this paper we completed the first step to this goal by discussing linear gravitational waves in de Sitter space-time.

Motivated by considerations of isolated systems discussed in section \ref{s2.1}, we focused on the upper Poincar\'e patch of de Sitter space-time. Isometries generated by 7 of the 10 de Sitter Killing fields leave this patch invariant. This group $\Gp$ is generated by 3 space-translations and 3 rotations that are tangential to the cosmological slices and a time translation that is transversal to them. Therefore, one expects well defined notions of linear and angular momentum, and energy, associated with any physical field on the Poincar\'e patch. We showed in section \ref{s2.2} that, in the case of Maxwell fields, these `conserved fluxes' arise as the Hamiltonians generating canonical transformations induced by the action of Killing fields on the covariant phase space $\psM$. Furthermore, in the $\Lambda=0$ case, the Hamiltonian framework has been used very effectively also for gravitational waves in \emph{full, nonlinear} general relativity: It leads to flux integrals corresponding to the Bondi-Metzner-Sachs (BMS) asymptotic symmetries \cite{aams}. Therefore, it is natural to use this strategy also in the $\Lambda >0$ case.

Since the covariant phase space consists of solutions to the field equations, in section \ref{s3} we discussed the asymptotic properties of solutions to linearized Einstein's equation in de Sitter space-time. In section \ref{s4} we constructed the covariant phase space $(\ps, \omega)$ of these linear gravitational waves. Each of the 7 Killing fields $K^{a}$ naturally defines flow on $\ps$ that preserves the symplectic structure $\omega$ thereon, and thus defines a Hamiltonian $\H_{K}$. These Hamiltonians provided us with the expressions (\ref{H3}) and (\ref{H4}) of fluxes of energy-momentum and angular momentum carried by gravitational waves. Furthermore, we could express these conserved fluxes in terms of fields defined on $\scrip$.

These results have a number of interesting features. First, to make the full nonlinear theory manageable in the $\Lambda >0$ case, at first it seems natural to strengthen the boundary conditions by requiring that the intrinsic geometry of $\scrip$ be conformally flat, as in  de Sitter space-time.  {However, almost 30 years ago Friedrich showed that the freely specifiable data at $\scri^{-}$ consists,
up to arbitrary conformal rescalings, of a freely specifiable Riemannian metric and a trace-free, symmetric tensor field of valence two, which satisfies a divergence equation \cite{hf}. Therefore, by applying those results to $\scrip$ (in place of $\scri^{-}$), it follows that demanding conformal flatness of the metric at $\scrip$ removes by hand part of this free data. In the linearized approximation, we could sharpen the implication of this condition. First, because the perturbed electric and magnetic parts of the Weyl curvature are gauge invariant, we can discuss the physical or true degrees of freedom, not just the freely specifiable data. Second, we could parametrize the gauge invariant content of a general linearized solution in terms of 4 functions $E^{(s)}$ and $B^{(s)}$ on $\scrip$ that capture these true (phase space) degrees of freedom. Finally, we showed that the additional condition at $\scrip$ sets $B^{(s)}=0$. Therefore, in the linear approximation one sees explicitly that this condition cuts the true degrees of freedom in gravitational waves exactly by half. Furthermore, the gravitational waves that do satisfy this condition carry no energy-momentum or angular momentum! Thus, although this strategy of gaining control over the nonlinear theory seems plausible at first, it is simply not viable. By isolating the true degrees of freedom at $\scrip$, it should be possible to show that this sharper results holds also in full general relativity with positive $\Lambda$.} 

Second, we found that the conserved energy has a peculiar feature: For matter fields as well as linearized gravity, energy $\H_{T}$ defined by the time translation $T^{a}$ can have either sign and, furthermore, is \emph{unbounded} below. Thus, there exist both electromagnetic and gravitational waves on de Sitter space-time which carry \emph{arbitrarily negative} energy, no matter \emph{how small} the positive $\Lambda$ is! This is in striking contrast with the $\Lambda=0$ situation, where the corresponding waves  carry strictly positive energy $\Ho_{t}$ in Minkowski space-times. How can one reconcile this strong contrast? What happens to the infinitely many solutions with large negative energy in the limit $\Lambda \to 0$? 

To analyze this issue, let us first recall that, to take this limit, one has to use the differential structure induced by the coordinates $(\vx,\, t)$. In this chart, the cosmological horizons which bound region I lie at  $r^{2} = (3/\Lambda) e^{-2Ht}$ (where $r^{2}=\vx\cdot\vx$). Therefore, in the limit $\Lambda \to 0$, region I in which $T^{a}$ is time-like fills out the whole Minkowski space. This is the geometric reason why even though $\H_{T}$ is unbounded below no matter how small the positive $\Lambda$ is, the limiting $\Ho_{t}$ is strictly positive. In the phase space language, as $\Lambda$ changes, the covariant phase space $\ps^{(\Lambda)}$, on which the Hamiltonian $\H_{T}$ are defined, \emph{itself changes.} In the limit, the set of solutions $h_{ab}$ on which $\H_{T}$ is negative simply disappears!

To summarize, as we showed explicitly in section \ref{s4.2.1}, there are families of metric perturbations $\gamma_{ab}(\Lambda)$ that satisfy the gauge conditions (\ref{gauge1}) and field equations (\ref{lee1}) for each $\Lambda$, and admit well-defined limits $\ho_{ab}$ as $\Lambda \to 0$ satisfying the standard gauge conditions and field equations (\ref{hoeqns}) in Minkowski space-time. This limiting procedure is onto: the limits $\ho_{ab}$ span the entire phase space $\pso$ of metric perturbations in Minkowski space. Furthermore along any of these families, the energy $\H_{T}\mid_{\gamma(\Lambda)}$  tends to the energy $\Ho_{T}\mid_{\ho}$ of the limiting perturbation in Minkowski space. Nonetheless, the lower bound of the energy function on phase spaces $\ps^{(\Lambda)}$ is discontinuous in the limit: It equals $-\infty$ for every $\Lambda$, however small, but vanishes for $\Lambda=0$.  

\begin{figure}[]
  \begin{center}
  \vskip-0.4cm
    %$a)$\hspace{8cm}$b)$
    \includegraphics[width=2.5in,height=2.6in,angle=0]{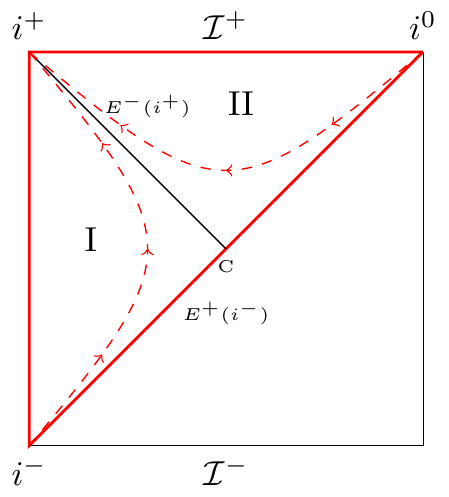}\hskip1.5cm
    \includegraphics[width=2.5in,height=2.6in,angle=0]{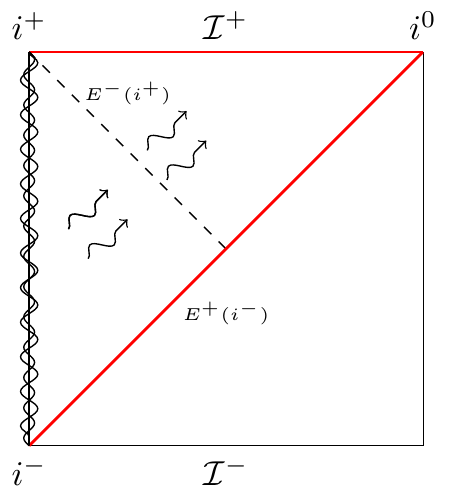}
\caption{\textit{Left Panel:} Integral curves of the time translation Killing field $T^{a}$ in the Poincar\'e patch. $T^{a}$ is future directed and time like in region I and space-like in region II. It is future directed and null on portion of the event horizon $E^{-}(i^{+})$ to the future of the cross-over 2-sphere (bifurcate horizon) $C$ and on the portion of the null event horizon $E^{+}(i^{-})$ to the past of $C$. It is \emph{past directed} and null on the portion of $E^{+}(i^{-})$ to the future of $C$. \textit{Right Panel:} A time changing quadrupole emitting gravitational waves. Radiation crossing the cosmological horizon $E^{-}(i^{+})$ and reaching $\scrip$ originates in region I; there is no incoming radiation on $E^{+}(i^{-})$.}
\label{fig2}
\end{center}
\end{figure}

Even though we do recover positivity of energy in the limit $\Lambda \to 0$, we are left with a conundrum because there is strong evidence that $\Lambda$ is small but non-zero in our universe: Can realistic gravitational waves have arbitrarily large negative energy in de Sitter space-time or, in the nonlinear context, in asymptotically de Sitter space-times? To probe this issue let us first analyze in some detail the origin of negative energy. Let us begin with Maxwell fields in de Sitter space-time. The stress-energy satisfies the dominant energy condition and the Killing field $T^{a}$ is future pointing on the part of $E^{+}(i^{-})$ that lies in region I and past pointing on the part that lies in region II  (see the left panel in Fig. \ref{fig2}). Therefore, the energy flux across $E^{+}(i^{-})$ into region I is positive \emph{but that into  region II is negative}. It is because of this negative flux into region II that the total energy can be negative. Therefore, if the Maxwell field under consideration vanished on the part of the horizon $E^{+}(i^{-})$ that lies in region II, the energy of those electromagnetic waves would be necessarily positive. For gravitational waves, we do not have a stress-energy tensor. However, using the fact that the Killing field $T^{a}$ is future directed and time-like in region I, it is easy to show that, if the initial data on any cosmological slice $\M$ were restricted to lie entirely in the intersection of $\M$ with region I,  the energy (\ref{H4}) of that cosmological perturbation is necessarily positive.% 
\footnote{This is most easily seen by using the symplectic form in (\ref{sym1}) to rewrite the energy as follows:
\be \H_{T} = -\, \f{1}{8\kappa H \eta} \,\int \rmd^{3}x\, \left( \left(\hat{r}^m - \eta^m \right) \Big(\hat{r}^n - \eta^n \right) + s^{mn} + 2 (1 + \f{r}{\eta}) \hat{r}^m \eta^n \Big) \,\mathring{\nabla}_m h_{ab} \mathring{\nabla}_n h_{cd} \, \, \qo^{ac}\qo^{bd} 
\ee
where $\qo_{ab} \hat{r}^a \hat{r}^b =1$ and $s^{mn} = \qo^{mn} - \hat{r}^m \hat{r}^n$. If the initial data is restricted to the intersection of $\M$ with region I we have $\left|r/ \eta\right| <1$ and, consequently, $\H_T$ is necessarily positive.}
In the limit $\eta \to -\infty$, the cosmological slice tends to $E^{+}(i^{-})$. Therefore, it again follows that the conserved energy flux at $\scrip$ can be negative only because there is a negative energy flux into the Poincar\'e patch across the part of $E^{+}(i^{-})$ that lies in region II. But in realistic situations gravitational waves from isolated systems would be generated entirely by a time changing quadrupole moment (depicted in the right panel of Fig. \ref{fig2}),   whence there would be no incoming flux across $E^{+}(i^{-})$ at all. The flux across $\scrip$ would just equal that across the future horizon $E^{-}(i^{+})$ that separates regions I and II. Since the Killing field $T^{a}$ is null and \emph{future directed} on this horizon, this flux has to be positive. Indeed we will show this explicitly in \cite{abk3}. Thus, in terms of fields \emph{at} $\scrip$, while general initial data \emph{can} have arbitrarily large negative energies, \emph{the initial data induced by gravitational waves produced by realistic sources is appropriately constrained for the energy flux across $\scrip$ to be positive.} In the linearized case, it appears to be rather straightforward to make these constraints explicit \cite{lh}. An interesting challenge in full nonlinear general relativity is to find the analogous constraints on fields at $\scrip$ induced by gravitational waves produced by realistic sources, in absence of incoming radiation \cite{ds4} (at least from the portion of the event horizon $E^{+}(i^{-})$ that lies to the future of the cross-over surface $C$). With these constraints at hand, one could hope to show that fluxes of energy carried by gravitational waves produced in physically realistic processes would be positive in full nonlinear general relativity with $\Lambda >0$, as one physically expects. 

{Finally, note that our entire analysis --and in particular the limit (\ref{mink-limit}) to Minkowski space-- was carried out by restricting ourselves to the future Poincar\'e patch of de Sitter space. As discussed in section \ref{s2.1}, in the description of isolated systems, this restriction is motivated by direct physical considerations. However, one may still ask if the results can be extended to full de Sitter space-time. The explicit form of the solutions we presented is indeed restricted to the future Poincar\'e patch because of the heavy use of the cosmological slicing. But each of these solutions admits a well-defined extension to full de Sitter space-time simply because every solution in our covariant phase space $\ps$ induces a well-defined initial data on the de Sitter Cauchy surfaces. For these extended solutions, our main results also hold on $\scri^{-}$. The central formula (\ref{H2}) holds for all ten Killing fields $K^{a}$ of de Sitter space-time in this extension.}

These constructions and results provide further guidance for the development of the gravitational radiation theory in full nonlinear general relativity with $\Lambda >0$. We will conclude this discussion with two examples. 

Consider first the problem of defining the 2-sphere energy-momentum and angular momentum charge integrals, analogous to the Bondi 4-momentum in the $\Lambda=0$ case. For a given prescription for selecting asymptotic symmetries, considerations involving field equations and geometry of $\scrip$  (discussed in section 5 of \cite{abk1}) suggest a natural, candidate expression for these charges for $\Lambda >0$. The difference between these integrals evaluated on any two 2-spheres on $\scrip$ provides a candidate expression of fluxes in the full theory across the region of $\scrip$ bounded by these 2-spheres. One can show that their linearization provides precisely the flux formulas (\ref{H3}) and (\ref{H5}) at $\scrip$, derived using completely independent Hamiltonian methods. This result provides a powerful hint for the charge integrals in the full nonlinear theory. The remaining open issue is the selection of appropriate asymptotic asymptotic symmetries, \emph{without assuming conformal flatness of the intrinsic geometry of $\scrip$} (which, as we showed, trivializes the situation by forcing all fluxes to vanish). 

A second issue in the full theory is the following. While observations strongly suggest that $\Lambda$ is positive in our universe, almost all analytical calculations and numerical simulations in gravitational wave science set $\Lambda$ to zero and work in the asymptotically Minkowskian context. (For notable exceptions, see \cite{num1,num2,num3}.) Since the actual value of $\Lambda$ is so small compared to the scales involved, say, in binary coalescences of astrophysical interest, it is natural to assume that setting $\Lambda$ to zero is an excellent approximation. However, it is not completely clear that this is true for two reasons. First, as we pointed out, the limit $\Lambda \to 0$ is discontinuous in important respects. Second, advanced LIGO will be eventually capable of detecting gravitational waves from sources that are $\sim 1$ Gpc away, a distance that is approximately 20\% of the cosmological radius. Therefore, apart from the intrinsic conceptual interest, it is important to be able to reliably calculate the `errors' one makes by setting $\Lambda$ to zero.%
\footnote{For example, there may be subtle effects --analogous to the nonlinear memory in the $\Lambda=0$ case-- that have remained under the radar so far.}
The details of the discussion of the $\Lambda  \to 0$ limit presented in this paper will help significantly in streamlining these calculations.

\section*{Acknowledgment}
 
We would like to thank Alejandro Corichi for discussions, Luis Lehner for correspondence and the referee for bringing to our attention Ref. 19. This work was supported in part by the NSF grant PHY-1205388, the Eberly research funds of Penn State and a Frymoyer Fellowship to AK and BB.

\end{document}